\documentclass[12pt]{report}
\usepackage[sort&compress]{natbib}% Example bibliographic style
\usepackage{NMTthesis2023}
\usepackage[hypertexnames=false]{hyperref}  % create live links in TOC
\hypersetup{
        colorlinks = true,
        linkcolor = black,
        anchorcolor = black,
        citecolor = black,
        filecolor = black,
        urlcolor = black,
	    pdfnewwindow = true,
        extension = pdf
} 
\mastersreport
\author{Leroy Jacob Valencia}

\title{Artificial Intelligence as the New Hacker: \protect 
Developing Agents for Offensive Security }
  
\degreeSought{Masters of Science in Transdiciplinary Cybersecurity}
 
\thesiscompletiondate{May 3, 2024}

\graduationdate{May, 2024}

\academicadvisor{Lorie Liebrock}
\researchadvisor{Danny Quist}
\committeesize{3}

\committeememberone{Jun Zheng}
\begin{document}
%
% Begin front matter
%
%----------------------------------------------------------------
%
% Produce the title page
%
\titlepage
%----------------------------------------------------------------
%----------------------------------------------------------------
%%%%\epigraph[who said it]{what they said}

%----------------------------------------------------------------
%%%%\frontispiece[title of the graphic]{some graphic}
%----------------------------------------------------------------

%----------------------------------------------------------------
\begin{dedication}
This work is dedicated to my wife, whose unwavering support and belief in my potential have been my constant source of strength and inspiration. 
\end{dedication}

%----------------------------------------------------------------
\begin{acknowledgments}
I wish to extend my sincerest gratitude to Danny Quist PhD., whose insights and guidance were invaluable throughout this research.  We also appreciate the support provided by New Mexico Cybersecurity Center of Excellence.  Lastly, I want to thank my peers and family members for their encouragement.
\end{acknowledgments}
%----------------------------------------------------------------
%
% An abstract is required.  
% After your abstract, provide two to six keywords, or key 
% phrases of up to three words, to assist librarians in
% indexing your work.
%
\begin{abstract}

In the vast domain of cybersecurity, the transition from reactive defense to offensive has become critical in protecting digital infrastructures. This paper explores the integration of Artificial Intelligence (AI) into offensive cybersecurity, particularly through the development of an autonomous AI agent, \textit{ReaperAI}, designed to simulate and execute cyberattacks. Leveraging the capabilities of Large Language Models (LLMs) such as GPT-4, ReaperAI demonstrates the potential to identify, exploit, and analyze security vulnerabilities autonomously.

This research outlines the core methodologies that can be utilized to increase consistency and performance, including task-driven penetration testing frameworks, AI-driven command generation, and advanced prompting techniques. The AI agent operates within a structured environment using Python, enhanced by Retrieval Augmented Generation (RAG) for contextual understanding and memory retention. ReaperAI was tested on platforms including, Hack The Box, where it successfully exploited known vulnerabilities, demonstrating its potential power.

However, the deployment of AI in offensive security presents significant ethical and operational challenges. The agent's development process revealed complexities in command execution, error handling, and maintaining ethical constraints, highlighting areas for future enhancement.

This study contributes to the discussion on AI's role in cybersecurity by showcasing how AI can augment offensive security strategies. It also proposes future research directions, including the refinement of AI interactions with cybersecurity tools, enhancement of learning mechanisms, and the discussion of ethical guidelines for AI in offensive roles. The findings advocate for a unique approach to AI implementation in cybersecurity, emphasizing innovation.

\end{abstract}

%----------------------------------------------------------------
%\signaturepage
\tableofcontents
%
% If you have no tables, comment out \listoftables.
%
\listoftables
%
% if you do not have figures, comment out the following lines
\listoffigures
%{
%\let\oldnumberline\numberline%
%\renewcommand{\numberline}{\figurename~\oldnumberline}
%\listoffigures
%}

%%%%\listofabbrs
% If you provide a List of Abbreviations, create a file
%   'abbrs.tex' containing the abbreviations table and
%    uncomment the following line
% 
%\listofabbrs
%
%----------------------------------------------------------------
\signaturepage
%----------------------------------------------------------------
%%%%\begin{preface}
%%%%
%%%%\end{preface}
%================================================================
% Body of the document.   Use one of these commands to start each
% chapter:
%   \chapter{FIRST CHAPTER TITLE}
%   \chapter[FIRST SHORT TITLE]{FULL LENGTH CHAPTER TITLE}
% If you have appendices, use these commands:
%   \appendix
%   \chapter{FIRST APPENDIX TITLE}
%     ...
%   \chapter{SECOND APPENDIX TITLE}
%     ...
%----------------------------------------------------------------
\chapter{Introduction}
In the rapidly changing domain of cybersecurity, defensive strategies have long been the focal point, emphasizing the protection of digital assets from malicious entities. However, with the increasing complexity and sophistication of cyber threats, the significance of offensive cybersecurity has grown, serving as an essential complement to traditional defensive tactics. Offensive cybersecurity employs the tactics, techniques, and mindset of adversaries to properly identify, exploit, and neutralize vulnerabilities before they can be exploited by attackers.

This paper examines the transformative role of Artificial Intelligence  in advancing offensive security measures. AI’s exceptional capabilities in processing extensive datasets, recognizing patterns, and automating intricate tasks make it a vital component in developing sophisticated offensive security strategies. This research provides an in-depth analysis of current technologies, methodologies, and the ethical concerns associated with AI in cybersecurity, highlighting AI’s potential.

Additionally, this paper discusses a GitHub project known as \textit{ReaperAI} \cite{ReaperAI}, which exemplifies the practical application of the concepts explored. \textit{ReaperAI} serves as a proof of concept, demonstrating how AI can integrate into offensive cybersecurity to effectively simulate an adversary. The aim of this research is to build upon existing AI studies, exploring new directions and the potential to harness existing AI capabilities to develop a functional product tailored for advanced cybersecurity solutions. This exploration not only contributes to the academic field but also to practical applications, pushing the boundaries of how AI can be leveraged in the context of offensive cybersecurity.

\section{Problem Statement}
This research seeks to answer the question "How can existing research on large language models be leveraged to develop a fully autonomous offensive security agent?". The objective is to develop a comprehensive compilation of research and ideas, drawing from future directions suggested in other scholarly articles and presentations. The effectiveness of the proposed methodologies and technologies will be evaluated based on the performance and behaviors exhibited by the agent's ability to produce a desired pentesting-like behavior.

\section{Impact}
The potential impact of this research on the industry could be significant. While there is ongoing speculation about the feasibility of such studies, there has yet to be a demonstration through a proof of concept or a minimum viable product executed in this manner. Successful implementation could set a precedent and inspire further innovation within the field.

\chapter{Background and Literature Review}
\section{Evolution of Offensive Security}
The world of offensive security has evolved quite a bit. It used to be mainly about simple vulnerability scanning, but now it involves advanced techniques that simulate actual cyberattacks. This shift brings to light the increasing complexity of cyber threats and how attackers are becoming more sophisticated by using more advanced techniques and tactics. In the past, security efforts were mostly about responding to incidents after they had happened. But now, with approaches like red teaming and ethical hacking, there's a focus on getting ahead of these issues by thinking like an attacker. This stance helps strengthen our defenses against cyber threats.

The early history of penetration testing dates back to the 1960s, when concerns about the security of computer communications were first raised. Government and businesses began forming teams to test and find vulnerabilities within their networks, to serve as a defense against any actual attacks. Notable contributions to this field were made by pioneers like James P. Anderson, who developed methodologies still in use today, like the Anderson Report \cite{infosecinstitute_history}.

Furthermore, the integration of continuous penetration testing with systems like a SIEM, security incident event management, are revolutionizing how vulnerabilities are identified and addressed, offering a more streamlined approach to cybersecurity. This integration helps in automating the response and remediation processes, thus enhancing the efficiency of security teams and reducing the time needed to address vulnerabilities \cite{guidepoint_security_2022}.

\section{Evolution of Large Language Models}
The evolution of Large Language Models  over the past decade illustrates a leap in artificial intelligence, transforming from basic natural language processing tools to highly sophisticated systems capable of generating human-like text and responses to human input. Initially, LLMs were limited in scope and capability in the early stages around 2018, focusing on specific tasks such as language translation and getting a response from a basic query. The significant publication of models like BERT (Bidirectional Encoder Representations from Transformers) in 2018 marked an advancement, introducing techniques that allowed for a deeper understanding of context within text generation \cite{devlin2018bert}. Subsequently, the release of GPT (Generative Pre-trained Transformer) by OpenAI further expanded the possibilities, employing unsupervised learning to generate coherent and contextually relevant text across a wide range of topics and formats \cite{radford2019language}. This progression from specialized applications to generalized capabilities reflects a broader trend in AI towards models that not only understand and generate text but also can be successful at conducting human like actions and exhibit nuanced understanding of complex subjects. This idea of understanding has begun a wave of advancement into the capabilities that are possible due to the sheer knowledge and data. The continuous growth in model size and sophistication, exemplified by the launch of GPT-3 \cite{openai_chatgpt} and its successors, really highlights an ongoing shift toward systems that can seamlessly integrate with human tasks and communication, prompting both ethical considerations and unique applications in technology and communication.

\section{AI in Cybersecurity: A Historical Perspective}

Historically, AI's involvement in cybersecurity has evolved from simple rule-based detection systems to more advanced machine learning algorithms that recognize complex patterns associated with cyber threats. "The evolving GenAI tools have been a double-edge sword in cybersecurity, benefiting both the defenders and the attackers." \cite{gupta2023chatgpt} AI's role has since broadened to encompass predictive analytics, automated response systems, and sophisticated threat intelligence that has mainly supported Blue Teams and Defense Teams. 

In the realm of offensive security, the use of AI has provided potential for changing traditional practices, in the simulation of realistic cyberattacks, automating the discovery of vulnerabilities and proof of exploitation, and examples of generating realistic phishing attacks. This integration of AI with offensive security tactics represents a strategy in cybersecurity, utilizing AI’s analytical capabilities to stay ahead of cybercriminals. \cite{Zennaro2023}

\section{Review of Current AI in Offensive Security}

Recent research in the field of cybersecurity has highlighted the significant role that artificial intelligence plays in enhancing offensive security measures. Advanced AI models, particularly those based on deep learning, are now relied on, in automating the detection of vulnerabilities, which was once manual and very labor-intensive. For example, the \textit{AutoPentest-DRL} framework employs deep reinforcement learning to automate and optimize penetration testing, allowing dynamic and efficient vulnerability exploitation within network systems \cite{autopentestdrl}. Moreover, AI's job of simulating complex cyberattack strategies through reinforcement learning models has expanded the scope of offensive security. Such models not only mimic attacker behaviors, but also innovate attack strategies, providing cybersecurity professionals with a tool set to anticipate potential breaches \cite{yang2022behaviourdiverse}. The integration of AI in penetration testing is also highlighted by its effectiveness in crafting sophisticated phishing emails that can evade standard detection systems, showcasing the ability of AI to adopt an attacker's mindset, which is truly alarming\cite{ermprotect2020ai}.  The adoption of large language models through natural language processing represents a particularly compelling advancement in artificial intelligence and machine learning due to the sheer level of intelligence, capable of understanding, reasoning, and offering suggestions and summarizations. This is made possible by their training on extensive internet datasets, making them highly valuable for complex tasks such as offensive security. In this paper, we will explore examples of these tools, as our discussion is primarily centered on the application of LLMs. 

\subsection{PentestGPT}
PentestGPT \cite{deng2023pentestgpt} is a sophisticated penetration testing tool that leverages the power of OpenAI's GPT-4 to automate and streamline the penetration testing process. Designed to function interactively, PentestGPT assists testers by guiding them through both the general progression of their penetration test and the execution of specific tasks. This tool is particularly adept at handling medium complexity Hack the Box machines and various Capture The Flag (CTF) challenges, enhancing the efficiency and precision of penetration tests.

The architecture of PentestGPT includes several modules that handle different aspects of the penetration testing workflow. It features a test generation module that generates the necessary commands for the testers to execute, a test reasoning module that aids in decision-making during the test, and a parsing module that interprets the outputs from the penetration tools and web interfaces. These components work together to provide a comprehensive and automated penetration testing solution.

PentestGPT has been shown to significantly outperform earlier models like GPT-3.5 in penetration testing tasks, achieving higher rates of task completion and demonstrating substantial improvements in operational efficiency. The development of PentestGPT reflects a notable advance in the use of LLMs for practical cybersecurity applications, offering a powerful tool that mimics the collaborative dynamics between experienced and novice testers in real-world settings.

\subsection{HackingbuddyGPT}
HackingbuddyGPT \cite{Happe_2023} is a cutting-edge tool designed to explore the potential of large language models in penetration testing, particularly focusing on Linux privilege escalation scenarios. Developed by the IPA Lab, hackingbuddyGPT integrates with OpenAI's GPT models to automate command generation for security testing. The tool operates by connecting via SSH to Linux targets (or SMB/PS Exec for Windows targets) and utilizes OpenAI's REST API-compatible models like GPT-3.5 Turbo and GPT-4 to suggest commands that could potentially expose vulnerabilities or escalate privileges. \cite{happe2023evaluating}

The system logs all run data, either into a file or in-memory, and features automatic root detection and a beautifully designed console output for better user interaction. One of the key functionalities of hackingbuddyGPT is its ability to limit rounds of interaction, which dictates how often the LLM will be queried for new commands, allowing for controlled testing scenarios. \cite{Happe_2023}

\section{LLM Limitations}
Large Language Models, such as GPT from OpenAI \cite{openai_chatgpt}, possess some serious capabilities in human natural language understanding and generation, but, they also encounter several limitations that can affect their functionality and integration into practical applications. Often, these limitations are misconstrued as signs of greater intelligence. In the following sections, we outline these limitations and explore their relevance to the field of penetration testing to ensure that we try and overcome them to produce an efficient workable proof of concept that isn't crippled by these limitations.

\subsection{Prompt Engineering}

Prompt engineering is a fairly new but critical aspect of using large-language models, that plays a significant role in specializing crafted inputs to steer the model toward improved quality generated outputs. This process is especially sensitive, as minor modifications in the prompt’s structure or phrasing of the prompt can lead to vastly different outcomes \cite{radford2019language}. Effective prompt engineering requires a deep understanding of the model's training data and embedded biases, which can be both labor-intensive and technically complex \cite{bender2021dangers}. This challenge typically involves considerable iterative adjustment and experimentation to refine interactions with the model and achieve optimal results \cite{liu2021pretrain}.

In the context of offensive cybersecurity, prompt engineering can significantly enhance the capabilities of an offensive agent. By precisely tailoring prompts, developers can direct the LLM to generate outputs that are more aligned with specific cybersecurity tasks, such as identifying vulnerabilities or providing commands to run on the terminal. This tailored approach allows for a more targeted and effective use of LLM in complex security environments, where generic responses may not suffice due to the sheer state size of one given problem in offensive security. Moreover, skilled prompt engineering can help mitigate the impact of biases in the model's responses, reducing the risk of generating inaccurate or harmful actions in sensitive security contexts as well as reduce any hallucinations the model might add that are not in factual data.

\subsection{Context \& Long Term Memory}

Context and memory present significant challenges in the effective deployment of large language models. Although LLMs handle brief segments of information effectively, their ability to retain or incorporate long-term context throughout a conversation or document is limited \cite{kagaya2024rap}. This limitation can lead to a deterioration of coherence in prolonged interactions, with the model possibly "forgetting" earlier segments of a conversation or struggling to sustain context across interactions \cite{wang2024memoryllm}. Commonly, remedies include integrating external systems to maintain the state or context, which can complicate the architecture of such systems and may adversely affect the accuracy and relevance of responses \cite{kagaya2024rap}. 

For instance, an offensive agent equipped with a supplementary memory system could better conduct and execute on prolonged penetration testing tasks, that require maintaining awareness of previous actions and their outcomes to become more effective. This integration introduces a more coherent and strategic approach to simulating or conducting cyberattacks to mimic a human actor who would adapt to dynamic targets or environments. While this complicates the architecture of LLM-based systems, the trade-off can lead to more robust and capable offensive tools in cybersecurity, where adaptability and persistence are crucial as the complexity grows.

\subsection{LLM Learning \& Reasoning}

Large language models, while proficient in language understanding due to their extensive pre-training on diverse datasets, do not adapt or learn dynamically post-deployment. Unlike some machine learning models that can continuously learn from new data, LLMs remain static unless they are retrained or fine-tuned with updated datasets. This characteristic restricts their utility in rapidly evolving fields without regular updates to their training material, which can be a resource-intensive process \cite{a16z2023}.

The challenge with LLMs is that they are not inherently equipped to integrate new information during operational use. Techniques such as prompt engineering are employed to mitigate this by carefully crafting inputs to guide the model's responses, yet this does not equate to learning from those interactions. To maintain relevance, particularly in dynamic fields, LLMs require periodic retraining or fine-tuning with new data, a process that demands both computational resources and expert oversight \cite{arizeai2023}.

However, this limitation can be mitigated by implementing periodic updates and fine-tuning sessions using the latest threat data, like CVEs, ensuring that the offensive agent remains updated with new tactics and vulnerabilities. Additionally, employing techniques like prompt engineering can help tailor the LLM’s output to simulate evolving attack scenarios more accurately, even within the constraints of its static knowledge base. This approach allows the offensive agent to remain a powerful tool in penetration testing, capable of adapting to new security landscapes through controlled updates rather than real-time learning, thus maintaining operational relevance and effectiveness.

\subsection{Command Parsing}

Command parsing with LLMs involves translating natural language commands into executable actions, which can be challenging due to the ambiguity and variability of human language. LLMs may misinterpret commands, especially those that are complex or have vague documentation, leading to incorrect or unsafe actions being conducted. Moreover, the ability of LLMs to understand context-dependent commands or those that require integration of multiple data sources is constrained by their training and the specific architectures used. This necessitates additional layers of validation and error handling in systems that rely on LLMs for command execution to ensure accuracy and safety in operations.

A method for transferring data not only between agents but also between code and large language models involves using JSON or a similar JSON-based standard. This approach facilitates the exchange of various types of information, including descriptions, outputs, and responses, ensuring a standardized communication format across different platforms and systems. 

\subsection{Training Data}
The limitations associated with training data significantly influence the development and effectiveness of large language models. The quality, diversity, and volume of the data used during training not only affect the model's performance but also its ability to function appropriately across various contexts. LLMs have the propensity to adopt and amplify biases from their training datasets, which can result in biased or detrimental output. Additionally, the reliance on vast datasets demands considerable computational resources, which can pose both environmental and economic challenges. Therefore, it is essential to ensure that the training data is both representative and ethically sourced to alleviate these issues.

When tailoring these models for specific domains, such as cybersecurity, introducing domain-specific knowledge is a complex endeavor. Experience with other tools has shown that, while more fine-tuned models can exhibit high degrees of specialization, they may also lack versatility if significant compromises are made during the fine-tuning process. Furthermore, as models increase in proficiency, their size tends to expand, often reaching hundreds of gigabytes. This increase in size can complicate deployment and operational efficiency. These considerations are particularly critical when developing an offensive agent, where the balance between model specificity, size, and adaptability must be carefully managed to ensure the production of a robust, production-grade agent.

\subsection{Risk/Fear}
Another limitation of Large Language Models  is their lack of inherent emotional capabilities, such as fear, which in humans plays a crucial role in risk assessment and decision-making. Humans often use fear as a heuristic for danger; it helps them avoid risks that could lead to harm. LLMs, in contrast, process decisions based on patterns and data without any emotional weighting. This can lead to challenges in situations where risk assessment is crucial, as models might not prioritize or evaluate threats effectively as a human would. This topic of study is fairly new as well and has not attracted much research in the context of advancing LLMs let alone using LLMs to expedite the field of risk analysis. \cite{esposito2024leveraging}

However, this limitation can also be viewed as an advantage, especially in the context of deploying LLMs as offensive agents in cybersecurity. The absence of fear allows LLMs to methodically execute tasks that would be considered high-risk or stressful for human operators. For instance, an LLM can engage in simulated cyber-attacks or test network vulnerabilities without hesitation or moral reservations, providing a thorough and relentless testing capability that might be compromised by human emotions.

\subsection{Creativity}
Lack of creativity is a major limitation of Large Language Models like GPT. While LLMs excel at generating content by recombining existing patterns and information from their vast training data, they do not truly "create" in the human sense of generating novel ideas from scratch. This limitation stems from the models' reliance on patterns and correlations within their training data, which confines their outputs to combinations of what has already been seen. Creativity, in contrast, often involves breaking away from established patterns to produce something genuinely new and original. There has been a framework developed recently that is used to benchmark this creativity to showcase limitations and provide potential for overcomings called: CreativeEval. \cite{delorenzo2024creativeval}

Creativity is a critical asset in cybersecurity, particularly for human penetration testers. These professionals thrive on their ability to think outside the box and devise innovative approaches to security testing, often breaking away from established patterns to uncover vulnerabilities that automated systems might overlook.

\subsection{Diligence}
Diligence in humans refers to their ability to consistently perform tasks accurately over time and to their fullest extent of knowledge and skills. Although LLMs can process and analyze large datasets with remarkable speed and accuracy, they lack the continuous attention to detail necessary for diligence. \cite{jin2024selfselected} They do not have the capability to self-assess their performance critically or improve independently without further training or updates, which can limit their application in environments requiring ongoing, meticulous attention to complex or changing data.

In the field of offensive security, the challenge of penetration testing is inherently complex, often demanding significant human investment in terms of time and expertise. While we can expect an LLM to deploy its full capabilities, it falls short in actively pursuing all possible avenues and may need to revisit information and tasks to improve outcomes, similar to what is required of human operators. LLMs have to be instructed and don't have natural knowledge of how the world works and especially how computers work.

\subsection{Situational Awareness}
LLMs typically lack situational awareness, which is critical in dynamic and context-dependent settings. They do not possess an understanding of the world in the same way humans do, nor can they interpret context beyond the scope of their training data. This limitation is particularly evident in scenarios requiring real-time decision-making or adaptation to new and unforeseen circumstances, which can impede their effectiveness in roles that require a high level of contextual adaptability. There have been systems that try to overcome this, like dynamic retrieval augmented generation where "Our framework is specifically designed to make decisions on when and what to retrieve based on the LLM's real-time information needs during the text generation process."  \cite{su2024dragin}

This limitation is a very steep hurdle for using LLMs to conduct offensive security due to the real time nature of the problem set that makes up penetration testing. Analyzing what is happening in real time is even a matter of exploitation or not with some time based attacks. Developing a way to properly integrate real time data into an LLMs prompts would be invaluable to not only offensive security but solving complex tasks all together.
 
\section{Agents}
GPT agents \cite{openai2023introducing}, or Generative Pre-trained Transformer agents, are also a very significant advancement in artificial intelligence and especially in LLMs. GPT models can be subsequently fine-tuned to perform specific tasks and are labeled as "agents".  These agents are built on the existing models that are initially pre-trained on a broad spectrum of internet-based text. This approach gives them the skills to grasp subtle context, generate text that's relevant and coherent, and handle language-based and domain specific tasks with impressive expertise.

The adaptability of GPT agents makes them applicable across a variety of fields, including customer service, content creation, summarization, programming, task automation, insight generation, and enhancing user interaction. Their capacity to continually learn from user engagements and adapt to novel information renders them invaluable for businesses aiming to harness AI to boost efficiency and engagement\cite{awesomegptagents}.

Further developments in GPT technology have introduced new methodologies where individual agents, each trained with a specific prompt, sequentially process and hand off their outcomes to subsequent agents. This model facilitates a more customized approach to problem-solving or response generation. During the period of this research, innovations like "Crew AI" \cite{crewai2023} and "AutoGenStudio" \cite{autogenstudio2023} emerged as leading platforms for creating such agent-based systems. This concept is particularly promising for developing autonomous systems like penetration testers, where the complexity of tasks necessitates decomposition into manageable, discrete segments that can be effectively handled by LLMs.

\chapter{Core Technologies and Methodologies}
The core technologies and methodologies employed in this project form the backbone of the research, emphasizing the integration of sophisticated AI tools with advanced cybersecurity practices. At the heart of the technology stack is \textit{gpt-4-turbo-preview} from OpenAI \cite{openai2024gpt4}, a state-of-the-art Large Language Model utilized for its expansive context understanding and dynamic response generation capabilities. This LLM serves as a central processing unit, driving the autonomous agents in decision-making and analytical tasks. Complementing the LLM, the Python wrapper plays a crucial role as the operational framework, managing interactions and ensuring seamless communication among various components of the system. Methodologically, the project adopts a hybrid approach that merges structured task trees with dynamic reprioritization capabilities, mirroring real-world penetration testing frameworks while incorporating the flexibility of AI-driven decision processes. This blend of cutting-edge AI technology and methodical security testing techniques ensures a comprehensive and adaptive system capable of addressing complex cybersecurity challenges in real-time.

\section{Pentesting Methodology}
\subsection{ATT\&CK Life Cycle}
The MITRE ATT\&CK  framework delineates a comprehensive catalog of tactics and techniques employed by cyber adversaries throughout the stages of a cyberattack \cite{mitre2024attack}. The initial phase, Reconnaissance, involves the systematic collection of data on potential targets. During this stage, attackers gather information to ascertain vulnerabilities and formulate an effective attack strategy. Methods employed include social engineering, network scanning, and the acquisition of publicly available data, which provide a broad understanding of the target’s defenses, technological infrastructure, and operational routines.

Following the reconnaissance stage is Vulnerability Analysis. In this phase, attackers analyze the accumulated information to pinpoint weaknesses within the target’s systems. The analysis typically involves the identification of security gaps such as outdated software components, system misconfigurations, and inadequate security policies. Advanced automated scanning tools may be deployed to detect these vulnerabilities, providing attackers with a clearer path for subsequent exploitation.

The final stage in the initial attack cycle is Exploitation. With vulnerabilities identified and strategies formulated, attackers exploit these weaknesses using various offensive measures. This stage involves the deployment of malware, use of exploit kits, and other intrusion techniques aimed at breaching security measures. The primary objective is to establish a secure foothold within the network, enabling further malicious activities such as data exfiltration, system compromise, or the dissemination of additional malicious payloads.

\begin{figure}
    \centering
    \includegraphics[width=1\linewidth]{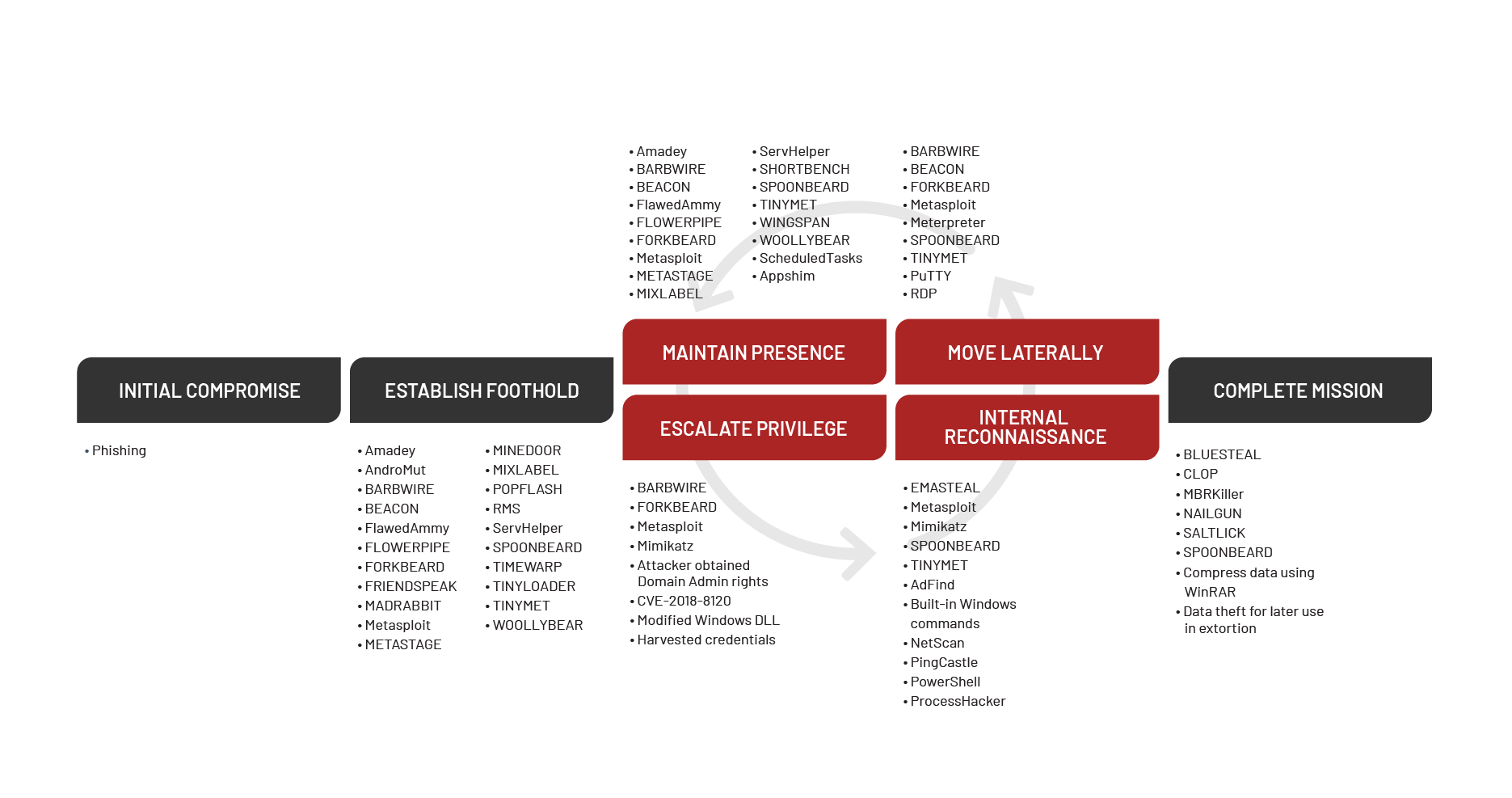}
    \caption{https://www.mandiant.com/resources/insights/targeted-attack-lifecycle}
    \label{fig:attck_life_cycle}
\end{figure}

A deep understanding of these stages is imperative for cybersecurity professionals. It aids in the formulation of robust defensive mechanisms designed to preemptively detect, thwart, and mitigate the actions of cyber adversaries before substantial damage is inflicted. The complete life cycle and be seen in Figure \ref{fig:attck_life_cycle}

\subsection{Hack The Box}
Hack The Box is an innovative online platform that provides a hands-on cybersecurity training environment for individuals and companies alike \cite{hackthebox2024}. It offers a variety of real-world scenarios through virtual labs, where users can practice hacking and test their penetration testing skills in a safe and legal setting. The platform features a range of challenges and machines that mimic different environments and security vulnerabilities, allowing users to engage in tasks ranging from simple puzzles to complex system exploits. Hack The Box also facilitates community interaction and learning, with forums and leaderboards that encourage competition and collaboration among users. This practical approach to learning cybersecurity is designed to sharpen problem-solving skills and provide real-time feedback, making it an invaluable resource for both aspiring and experienced cybersecurity professionals looking to enhance their offensive security capabilities. 

Hack The Box is a common test bed for implementing and refining penetration testing methodologies. ReaperAI orchestrates tasks that mimic real penetration test scenarios, including reconnaissance, vulnerability assessment, exploitation, and post-exploitation, all crucial elements in HTB machines. HTB provides a structured yet adaptable platform that allows the research to be applied and tested, proving more advantageous than self-hosted VM setups due to ease of use and setup efficiency.

\section{Methodology}

\begin{figure}
    \centering
    \includegraphics[width=1\linewidth]{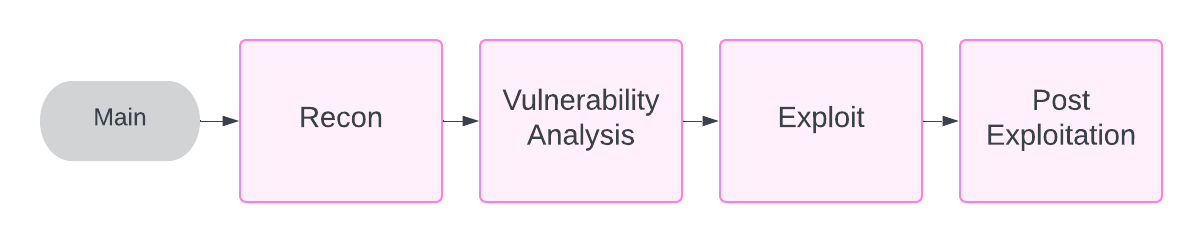}
    \caption{Main Logic Flow}
    \label{fig:enter-label}
\end{figure}
\subsection{Integration with LLM}

For this research, the decision was made to employ \textit{gpt-turbo-4-preview} with a substantial context size of 128k tokens, reflecting the most sophisticated technology available at the time of testing. The primary aim of this choice was not to compare various Large Language Models, but rather to explore and demonstrate the advanced capabilities highlighted in existing foundational research. As of April 22, 2024, the rate per one million tokens worth of input is \$10 and \$30 per one million output. This study in total cost about \$40 dollars in research, development, and testing. This approach ensures that the study focuses on leveraging, at the time, cutting-edge AI capabilities to assess their practical applications and effectiveness in complex computational tasks.

\subsection{Autonomous Agents}

This project aimed to develop a fully autonomous agent that could operate independently without human intervention. This objective brings forth its own set of challenges and complexities, particularly in ensuring robust and reliable operations while balancing cutting edge technology. The Python wrapper used in this configuration serves as the central nervous system, orchestrating interactions and maintaining seamless communication between agent loops, LLMs, terminals, and python code. These LLM agents are tasked with autonomous reasoning and decision-making that simulate a high level of cognitive processing akin to human-like thinking and problem-solving skills to ensure that the agent can act as if is a human operator.

\subsection{Objectives and Tasks}

Unlike the approach taken in hackingbuddyGPT, which follows the BabyAGI \cite{nakajima2024babyagi} model of executing an action within a task, enriching it with context, and then reassessing priorities, this project introduces a structured task tree methodology. While penetration testing typically follows a systematic approach similar to a task tree, it is crucial to incorporate the dynamic element of BabyAGI, where new and critical information can prompt immediate reprioritization and strategic shifts. This dual approach ensures that the agents not only adhere to a structured methodology but also remains flexible and responsive to new insights and challenges.

\subsection{Decision-Making in Tasks}

The autonomous decision-making process in these agents was crafted around the completion of specific tasks. It was crucial to design a system that was not limited to pre-defined, hard-coded strategies, but instead could adapt based on situational demands. To achieve this flexibility, the developed evaluation method incorporates concepts of diminishing returns and strict time constraints—both reflective of the nature required in penetration testing. This system evaluates task completion through dynamically generated prompts, which assess whether the tasks have been accomplished based on historical data and prior analyses.

\subsection{Analyzation}

Building upon the initial concepts of analyzation introduced by \textit{hackingbuddyGPT}, the methodology was enhanced to provide a more sophisticated analysis of actions and outcomes. This upgraded approach not only assesses what has transpired, but also generates  recommendations for subsequent steps. Such ongoing analyzation is crucial for the continuous improvement and adaptation of the agents, ensuring they remain effective and relevant as they interact with complex environments.

\subsection{Evaluation}

This phase of the project introduces a new task evaluation concept that was not covered by \textit{hackingbuddyGPT}, inspired by traditional human-led decision-making processes. In this program, an agent performs an action, then evaluates the results to determine if they suffice for the task at hand—or if only progress towards the task has been made. This evaluative process is critical and mirrors the feedback mechanism in reinforcement learning models, where the agent learns and adapts based on success feedback.

\chapter{Building the AI Offensive Agent}

The design of the AI offensive agent is rooted in the principle of leveraging the LLM technology described previously as well as the methodology described previously to simulate and understand offensive cybersecurity tasks. The heart of this system is the LLM, accessible through Python classes and APIs, while a Python wrapper serves as the core logic driver to process functions and essentially be the body of the LLM. This architecture allows for innovated integration of advanced AI capabilities with current cybersecurity tools and frameworks, providing a unique concept for simulating cyber-attacks, analyzing potential vulnerabilities, and automating the decision-making process. The Python wrapper facilitates easy access to the LLM's functionalities, enabling the dynamic construction of queries and the interpretation of responses for further processing. The core for this development was based on the foundation of \textit{hackingbuddyGPT} which was enhanced to become \textit{ReaperAI}, a proof of concept fully autonomous offensive agent.  \cite{Happe_2023}

\section{Agents}
\begin{figure}
    \centering
    \includegraphics[width=1\linewidth]{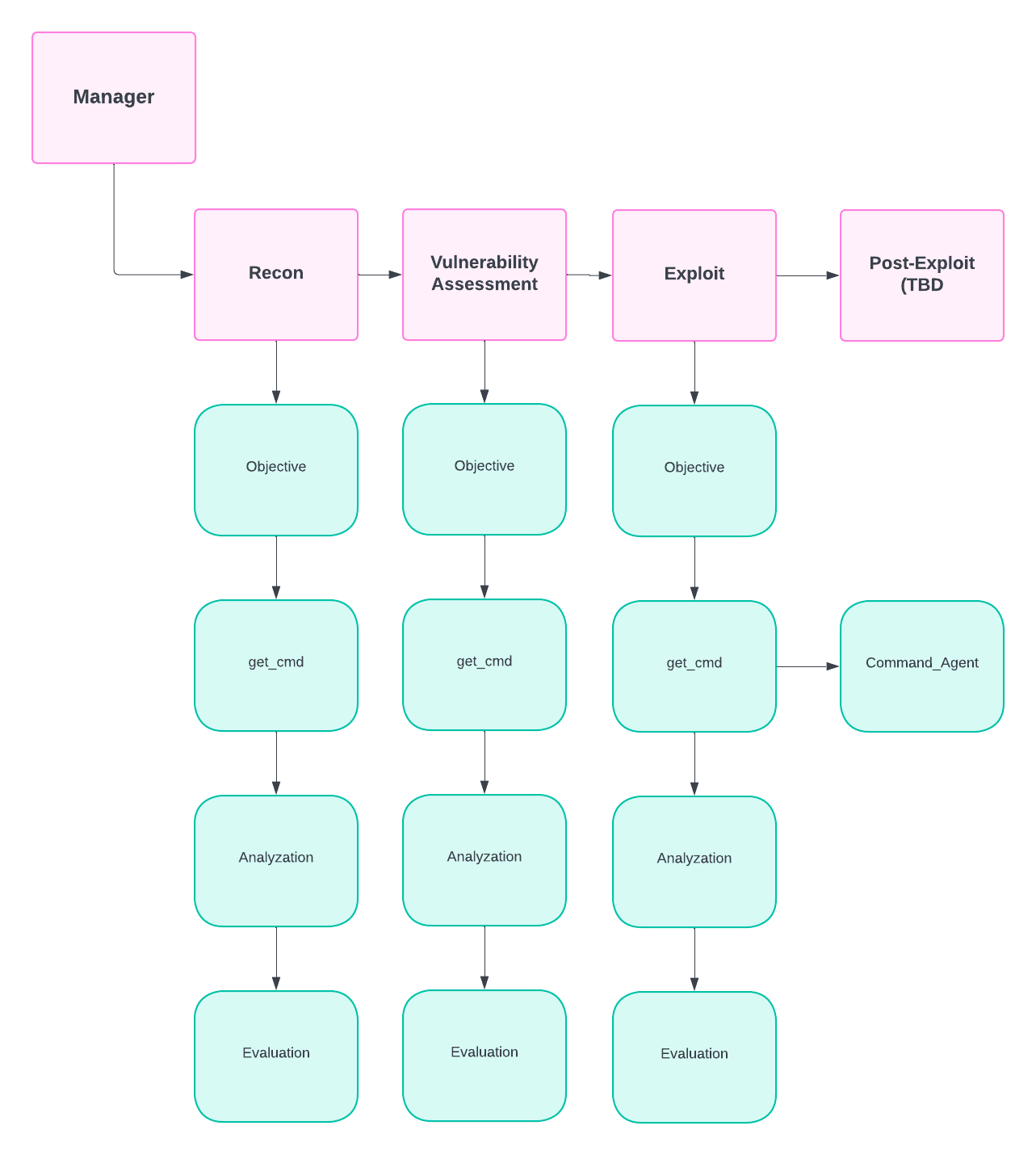}
    \caption{Agent Hierarchy}
    \label{fig:agent-hierarchy}
\end{figure}

The design approach for this agent involves using a subset of specialized sub-agents, each tasked with executing more narrowly defined functions to enhance result quality while ensuring communication upstream to the parent agent. Merely giving the LLM the task to \textit{"complete a pentest"} is too vast and too vague. The only predefined hard coded workflow was the pentesting methodology to prevent the AI from deviating and ensure it remains focused on its intended purpose, rather than determining its own fundamental objectives of a penetration test. Previous experimentation showed that the LLMs knowledge set did contain "\textit{steps to complete a black box penetration test}" but did not produce consistent and accurate fundamental objectives to be allowed to generate them autonomously. The high-level agent tree can be seen in \ref{fig:agent-hierarchy}

\section{Prompting, Decision Making, and Natural Language Understanding}

\subsection{Templating}
In Happe's project, implementation of the Mako templating library, was a good foundation but was ultimately needed to be expanded to achieve the level of quality required for this study \cite{makotemplates}. Each prompt functions as a sub-agent within a "prompt chaining" approach \cite{PromptChaining2023}. The prompts themselves are stored in .txt files, which contain templating text. This setup allows the text files to act as variables where various inputs can be introduced. For instance, state history, commands, and other contextual data can be inserted to enhance the quality of the interactions. This approach integrates several prompting techniques, including few-shot learning and chains of prompting, among others, to ensure effective and efficient performance.

\subsection{Enhanced Decision-Making Through Natural Language Prompting}

At the heart of the agent's functionality is its sophisticated use of natural language prompting to guide decision-making and reasoning processes. This method involves framing cybersecurity tasks within natural language prompts that the Large Language Model processes. By utilizing the LLM's advanced language comprehension abilities, the system can generate insights, strategies, and responses that closely resemble the thought processes of experienced human security experts. This strategic use of language-based prompts enhances the agent's ability to reason and decide on the most effective course of action in complex security situations. Through this, ReaperAI has a combined way of prompting techniques that at the time of research are fairly new. This integrates concepts like Role Prompting, Chain-of-Prompting, Chain-of-Thought, Real-Time prompt optimization. These are all fairly new techniques that were described earlier and all combined to create the prompts in ReaperAI as seen in Table \ref{tab:prompt-engineering}

\begin{table}[hbt!]
    \centering
    \begin{tabular}{p{0.25\linewidth} | p{0.6\linewidth}}
        \textbf{ Technique}& \textbf{Intent}\\ \hline
         Role Prompting& Bypass filters that would be used in the generic role\\
         Chain-of-Prompting& Chain prompts together to allow a bigger task to be fullfilled\\
         Chain-of-Thought& Chain the thoughts together on a pentest to make a decision\\
         Real-Time prompt optimization& Provide real-time information to the LLM\\\
    \end{tabular}
    \caption{Table of Intent for Prompting Techniques}
    \label{tab:prompt-engineering}
\end{table}

\subsection{Adaptive Decision-Making}

Adaptive decision-making is a core feature of ReaperAI, allowing it to dynamically adjust its strategies based on the analysis of command outputs and the current state of the system. This flexibility is crucial for navigating the complex landscape of penetration testing, where conditions can change unpredictably. By evaluating the effectiveness of each command and its impact on the system, ReaperAI can decide whether to alter command sequences, repeat commands, or adjust arguments according to the recommendations provided by the other LLM agents who analyze output. This adaptive approach ensures that the testing strategy remains aligned with the evolving security environment, maximizing the effectiveness of the test and ensuring that all security vulnerabilities are thoroughly explored and addressed. The ability to integrate new insights helps maintain the relevance and efficacy of the penetration testing process, ensuring that each action taken is informed by the most current data and expert system analysis via the prompt injection.

\subsection{Minimizing Unwanted Behaviors Through Precise Prompt Engineering}

To prevent unwanted behaviors such as irrelevant command outputs or overly detailed explanations, the system employs the \textit{Mako} templating engine described above. This engine integrates data from the ReaperAI’s Python logic into the prompts, which are then passed to a prompt creation function.  This approach minimizes the need for extensive prompt engineering by streamlining the interaction with the LLM, focusing mainly on crafting basic, targeted inquiries that enhance the quality of the generated responses. By manipulating prompts to exclude undesired outputs, as seen in Table \ref{tab:prompt-cleaning}, the system maintains preciseness in its operations, which result in a more effective and efficient problem-solving capabilities within the scope.

\begin{table}[hbt!]
    \centering

    \begin{tabular}{|l|} \hline 
        \textbf{Prompt Cleaning}\\ \hline 
        "explain step by step"\\ \hline 
        "Provide a recommendation."\\ \hline 
        "Output the list in a json array."\\ \hline 
        "Do NOT give any
explanations or descriptions."\\ \hline 
 "Do not
include any explanations or any prefixes. Only provide the command to run.."\\ \hline 
 "As well as serve
for in context memory."\\ \hline
        \end{tabular}
\caption{Prompts that aided in Cleaning Response}
\label{tab:prompt-cleaning}

\end{table}

% Make a reference to appendixes where example prompts are located

\section{RAG for Enhanced Memory and Contextual Understanding}

\subsection{PostgreSQL + Python Classes}
To tackle the challenges of memory retention and contextual awareness in AI-driven cybersecurity tasks, the system employs a Retrieval Augmented Generation (RAG) component. This innovative approach leverages the extensive knowledge base of pre-trained models, enhancing it with the capability to retrieve relevant information and generate responses that are contextually aware. The integration of RAG significantly boosts the agent's ability to remember prior interactions, comprehend complex command sequences, and make decisions informed by historical data and recognized patterns. This enhancement is pivotal for maintaining a continuous state across individual and subsequent sessions, ensuring that the agent can seamlessly resume its tasks without losing context.

ReaperAI utilizes this RAG capability by interacting with a database to access necessary information according to the agent’s operational functions. For instance, when generating the next command, the system retrieves the current state and analyses of the previous command from the database to improve the quality and relevance of the forthcoming command outputs. This cycle of retrieval and generation ensures that each decision is as informed as possible.

However, the application of RAG in ReaperAI deviates from traditional uses, particularly due to the unique domain of offensive security. The cybersecurity domain lacks the vector text and object embeddings required for conventional vector databases, which poses a significant challenge. This specific issue and potential solutions will be further discussed in the "Future Directions" section of this paper, highlighting the ongoing adaptation of RAG technology to meet specialized cybersecurity needs.

To assist the agent in managing tasks and tracking hosts, Python classes were implemented. These classes are standard Python constructs that were designed to process information generated by the Large Language Model  and to feed this information back into the LLM as needed. This cyclical interaction helps to structure the tasks efficiently, allowing the agent to maintain a clear and organized workflow. In this approach, Python classes are utilized as intermediaries to encapsulate task and host details, which simplifies the management of complex data and interactions. This structured framework not only enhances code clarity and maintainability but also empowers the Large Language Model to effectively manage and update task-specific and host-specific information. Consequently, this ensures all operations are coherent and well-aligned.

\section{Task-Driven Methodology}

\subsection{Task Tree Management}

\begin{figure}[hbt!] 
    \centering
    \includegraphics[width=1\linewidth]{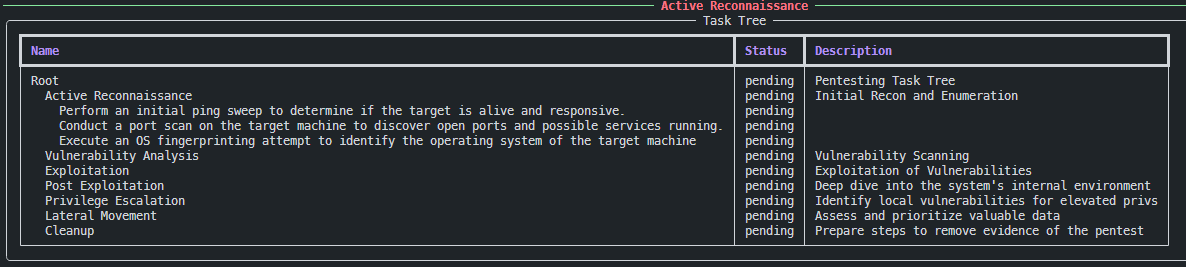}
    \caption{Example Task Tree}
    \label{fig:example-task-tree}
\end{figure}

In the realm of penetration testing, the ReaperAI introduces a methodical structure through the implementation of a task tree, which separates the entire process into distinct stages: reconnaissance, vulnerability analysis, exploitation. This organizational strategy is vital for systematically managing the complex procedures involved in penetration testing. The LLM will be tasked to breaking down the process into manageable sections and loading those results into the task tree to ensure that each phase is executed and is in alignment with the overall testing strategy. It facilitates easier monitoring and progression through tasks, allowing focus on one segment at a time while maintaining an overview of the entire testing landscape. This structured approach not only streamlines the testing process but also enhances the effectiveness of the tests by ensuring thorough coverage of all necessary aspects of the system’s security. It is worth mentioning that in Figure \ref{fig:example-task-tree} there are only 3 sub-tasks under "Active Reconnaissance" this is specifically due to the fact that when generating more than 3 sub-tasks the LLM would start to assume information about the system that was not actually true. For example, one task would be to "Use gobuster to enumerate web directories" but the LLM hasn't actually conducted any scans to know that is a valid task. To patch the solution of task validation, reducing the number of tasks provided better results in staying consistent without demonstrating tasks that had information that was assumed.

\subsection{Dynamic Task Updates}

\begin{figure}[hbt!] 
    \centering
    \includegraphics[width=1\linewidth]{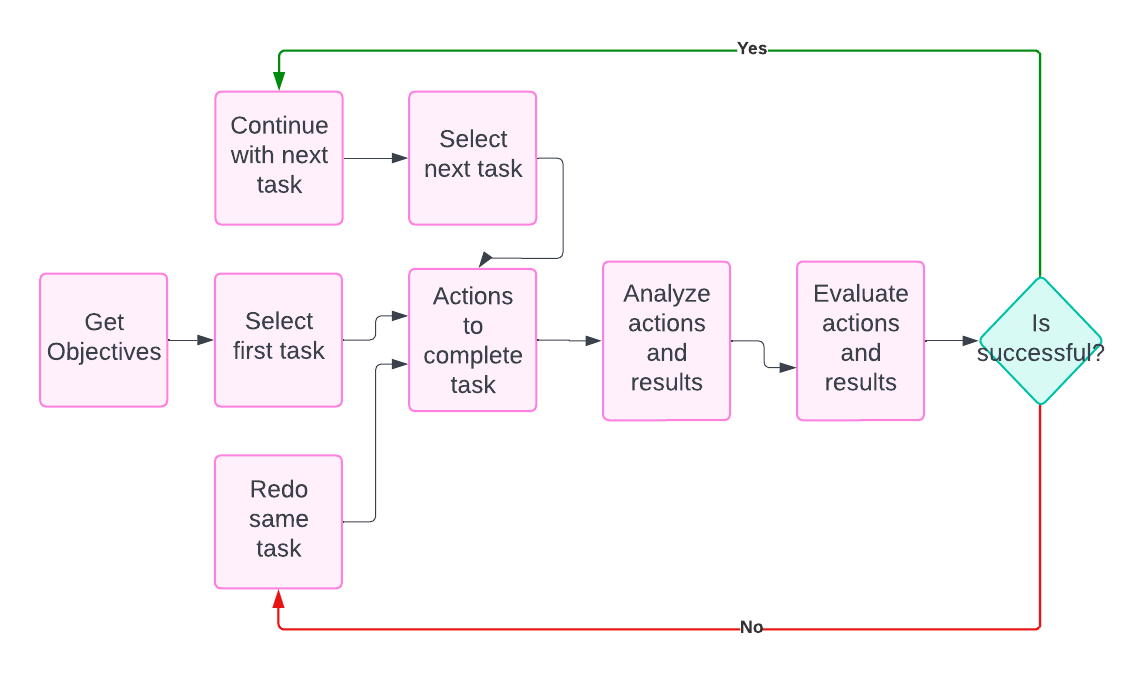}
    \caption{Decision Making on Tasks}
    \label{fig:main-decision-flow}
\end{figure}

The dynamic nature of security environments requires an equally agile response during penetration testing, which the ReaperAI addresses through real-time updates to tasks based on outcomes and feedback from the Large Language Model. As the penetration testing progresses, each action's result is analyzed and the subsequent tasks are generated based on the previous collection of information. This adaptive method allows the testing process to remain flexible and responsive, accommodating changes and unexpected results as they occur. For instance, if an expected vulnerability is not found, the task tree won't include any of the vulnerabilities-centric paths in order to prevent tangential rabbit holes. Similarly, successful exploitation might lead to additional tasks focusing on deeper system analysis or cleanup. This real-time feedback loop ensures that the penetration testing is not only thorough but also maximally efficient, adapting on-the-fly to findings and shifting priorities without losing momentum. The actual workflow is represented in Figure\ref{fig:main-decision-flow}.

\section{AI-Driven Command Generation and Processing}

\subsection{Workflow}
\begin{figure}[hbt!]
    \centering
    \includegraphics[width=1\linewidth]{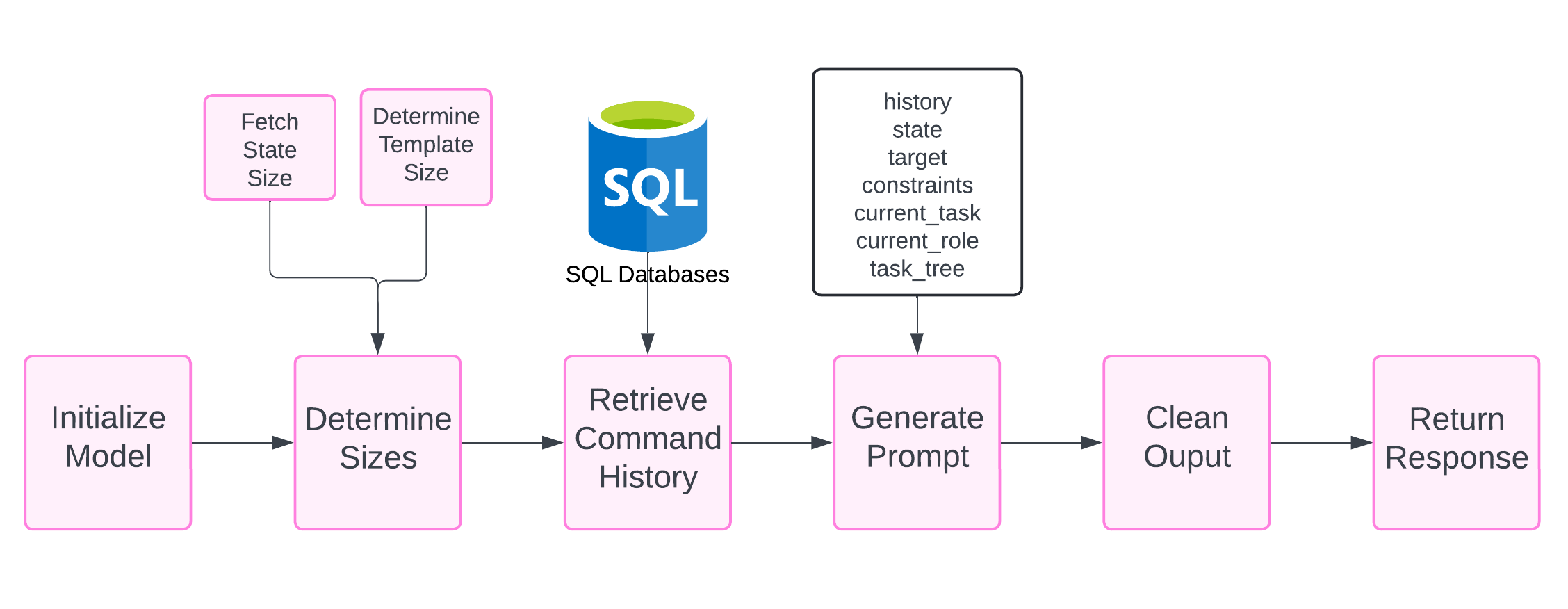}
    \caption{Get Next Command Workflow}
    \label{fig:get_next_command}
\end{figure}

In Figure \ref{fig:get_next_command} which shows the workflow for the main command generation process of the agent. The process begins with initializing the program by connecting to and retrieving it from the LLM. Next, the size determinations are made by fetching the current state size using \textbf{get\_state\_size} and determining the template size with \textbf{num\_tokens\_from\_string}, based on a source template to ensure token requirements aren't being exceeded. Following size determinations, the command history is retrieved through the \textbf{get\_cmd\_history\_v3} function, which combines the state size, template size, and other relevant parameters from the database and memory of the wrapper. A text prompt for the LLM is then generated using \textbf{create\_and\_ask\_prompt\_text}, incorporating all necessary parameters such as history, state, target, constraints, current task, current role, task tree, and details of the analysis. Once the LLM has processed the prompt, the output is cleaned using \textbf{command\_output\_cleaner} to ensure the response doesn't have residual artifacts included by the LLM like \textbf{\$} or \textbf{ bash}. Finally, the process concludes with returning the cleaned response, completing the interaction cycle with the LLM.

\subsection{Integration with LLM}

The program capitalizes on the advanced capabilities of LLMs by establishing a connection to an LLM server, using an API key. There are financial costs to using OpenAI's API, which are priced by the million tokens due to its closed source subscription model. These costs are not too expensive, but are important factors to consider when discussing the capabilities of an offensive agent. This integration is crucial, however, as it harnesses the AI's ability to generate and process commands based on vast datasets it was trained on. By utilizing AI to generate actionable commands and interpret outputs through the use of common communication protocol, REST API, the program reduces the manual effort required in formulating commands and speeds up the testing process. This automation not only increases the autonomous nature of the tests but also enhances their accuracy by leveraging the LLM’s easily available communication interface. The AI's input helps ensure that the commands are both contextually relevant and highly optimized for the tasks at hand, thereby streamlining the workflow during penetration testing.

\begin{figure}[hbt!] 
    \centering
    \includegraphics[width=1\linewidth]{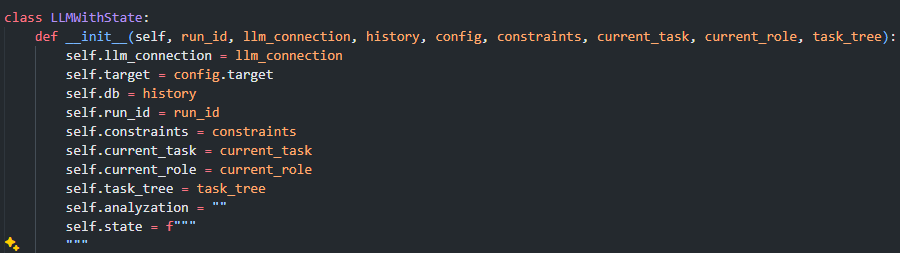}
    \caption{LLM Object}
    \label{fig:llm-object}
\end{figure}

In \textit{ReaperAI}, the idea from Happe's project was to create a class for the LLM to ensure that the state and other functions and constants would stay contained in it shown in Figure\ref{fig:llm-object}  This implementation of standard class/object behavior, common in most programming languages, was chosen as the most suitable for the desired functionality of the LLM.

\subsection{Stateful Interaction}

To ensure the continuity and relevance of interactions within the dynamic environment of penetration testing, the script maintains a stateful interaction with the LLM. This approach helps preserve the context of the penetration test across different interactions with the system, a critical aspect for maintaining the accuracy and relevance of AI-generated suggestions. By keeping track of previous commands and responses, the stateful system can provide contextually appropriate suggestions that build on earlier actions, thereby avoiding redundant or irrelevant commands. 

\begin{figure}[hbt!]
    \centering
    \includegraphics[width=1\linewidth]{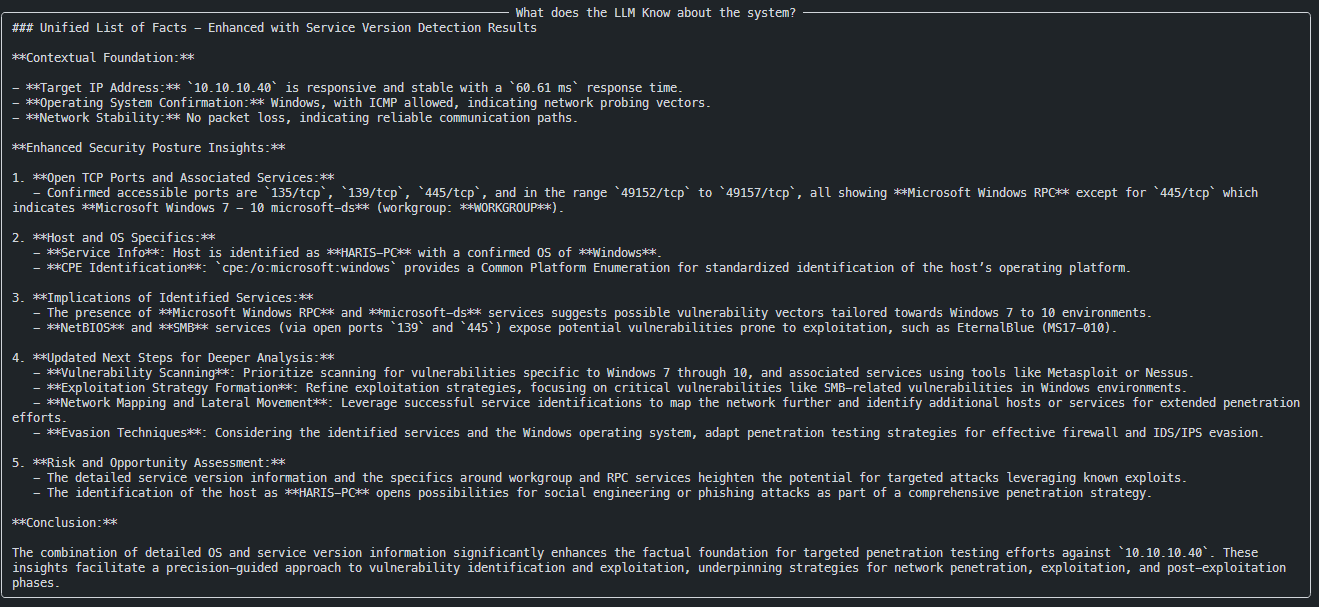}
    \caption{A Sample of State at a Given Time}
    \label{fig:example-state}
\end{figure}

Analysis is crucial both for the large language model  and the human overseeing it. By providing a summary at each main step of the workflow, as seen in Figure\ref{fig:example-state}, the LLM is equipped to reflect on its recent actions and respond appropriately. This level of analysis is also beneficial for the human observer, enabling them to monitor the LLM's performance and ensure that it is operating correctly. This dual focus on analysis helps maintain the integrity and effectiveness of the process.

\begin{figure}[hbt!]
    \centering
    \includegraphics[width=1\linewidth]{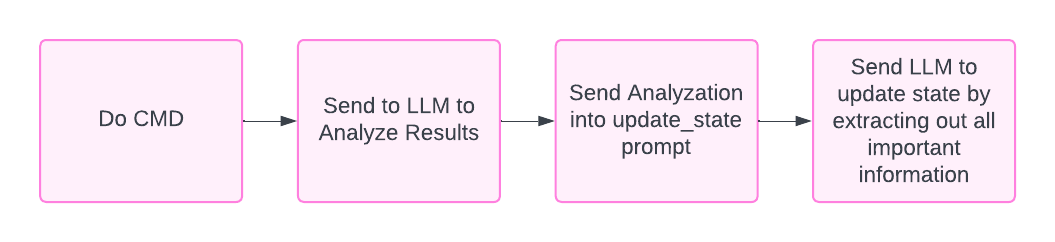}
    \caption{ State Workflow}
    \label{fig:state-workflow}
\end{figure}

\begin{figure}[hbt!]
    \centering
    \includegraphics[width=1\linewidth]{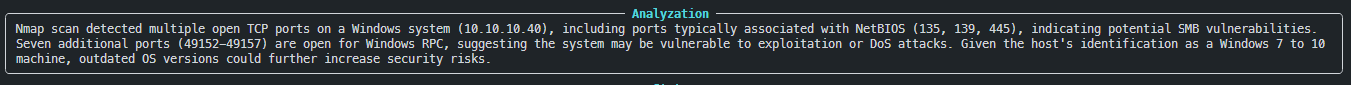}
    \caption{Sample Analyzation at a Given Time}
    \label{fig:sample-analyzation}
\end{figure}

This method, seen in Figure \ref{fig:state-workflow}, is essential for conducting comprehensive and effective penetration tests, as it allows the AI to adapt its recommendations based on real-time data and the evolving state of the system being tested. This ongoing contextual awareness, seen in Figure \ref{fig:sample-analyzation}, ensures that AI’s contributions are not only technically appropriate but also strategically astute, thereby enhancing the overall effectiveness of the penetration testing process by also giving a perspective view on what was just conducted on the terminal.

\section{Command Execution}
\subsection{Non-Interactive Execution}
Arguably, one of the most complex aspects of this project involved devising a unique method for the large language model to interact with a program. This paper previously outlined the significant challenge of lacking a standardized approach for establishing bidirectional communication between a Python program and the LLM. In the \textit{ReaperAI} system, JSON and structured prompts serve as the main channels for this interaction, ensuring that outputs from the LLM are consistent, and well-formatted to allow parsing from within Python. Although the LLM can process a broad spectrum of information, the primary difficulty resides in parsing, extracting, and applying the right information from the LLM and using that in a way that is effective.

This execution strategy draws inspiration from the concepts presented by Happe and hackingbuddyGPT \cite{Happe_2023} in 2023, yet deviates from their model by not using SSH to execute commands remotely. Instead, commands are run locally on a Kali machine using Python's \textit{subprocess} piping mechanism. Depending on the objectives and current tasks, ReaperAI formats this information into a prompt to solicit a command from the LLM. The command received from the LLM, structured as a JSON output, is then converted into an actual command string that the \textit{subprocess} can execute. This method ensures a seamless translation of LLM outputs into executable actions, optimizing the interaction between the LLM and the Python environment.

\subsection{Interactive Execution}
\begin{figure}[hbt!]
    \centering
    \includegraphics[width=1\linewidth]{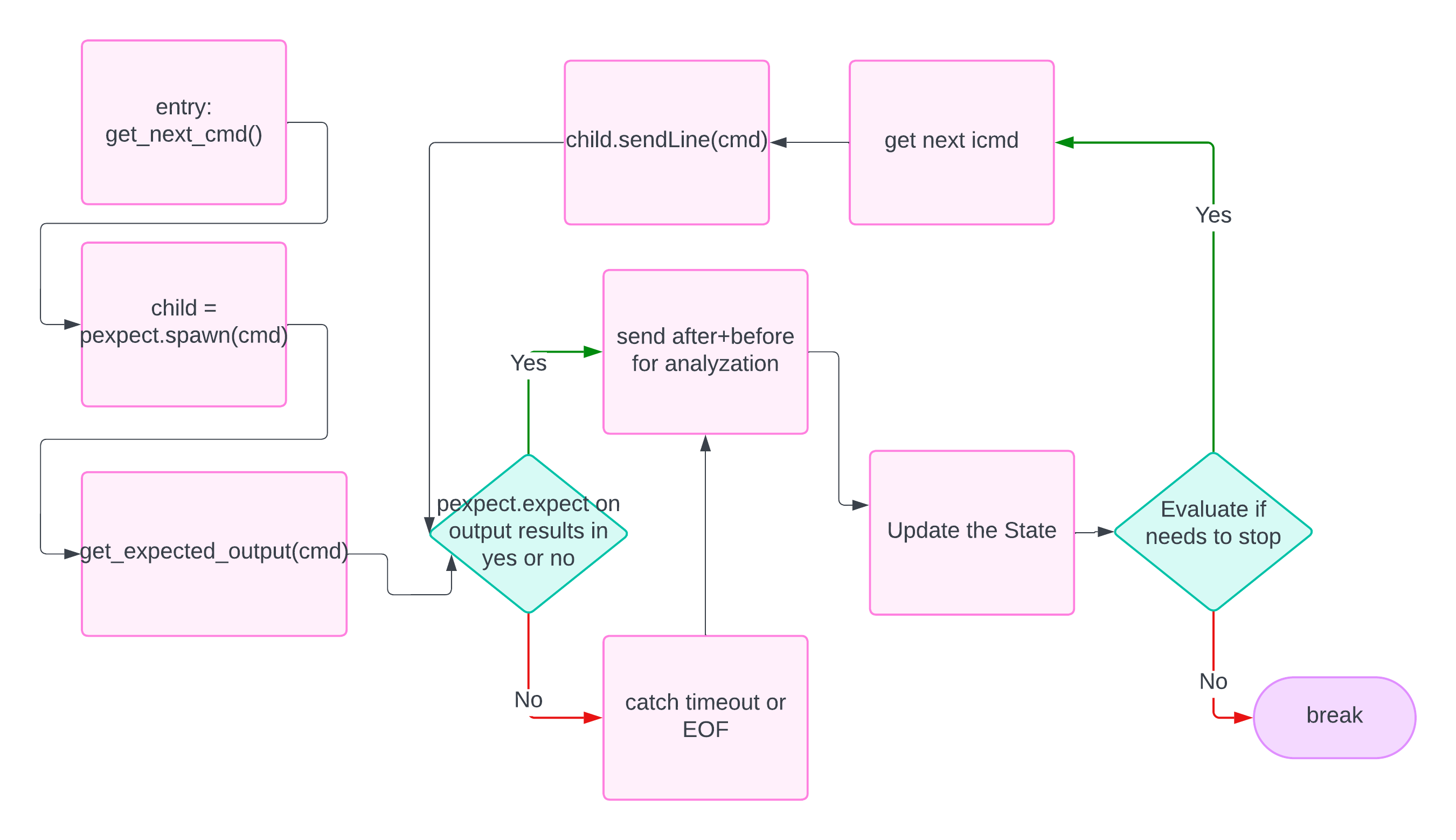}
    \caption{Interactive Command Workflow}
    \label{fig:interactive-commands}
\end{figure}

Interactive execution is also a crucial feature of the script, facilitated by a tool called '\textit{pexpect}' \cite{pexpect}. This tool allows the agent to interact dynamically with the command-line interface, handling commands generated by the LLM. The reason for the attention to this is that traditional one time run programs signal an end to the terminal with an EOF, so the operator knows when to read the stdout. When commands prompt for user input, the EOF has not been reached yet, so we have to resort to another library for this concept. "\textit{Pexpect is a pure Python module for spawning child applications; controlling them; and responding to expected patterns in their output. Pexpect works like Don Libes’ Expect. Pexpect allows your script to spawn a child application and control it as if a human were typing commands}" \cite{pexpect}. This simulates a human-like interaction with the system. This process is managed by a separate command agent, which determines the appropriate times to send new inputs or read outputs from the command line, enhancing the program's ability to handle complex sequences of commands that require interactive responses. This functionality is still not fully supported in \textit{ReaperAI}, but can be seen in Figure \ref{fig:interactive-commands}. The proof of concept demoed in \textit{ReaperAI}, is based around metasploit, but the workflow was designed to be universal towards other interactive programs like smbclient, netcat, etc. 

\subsubsection{Reading of Interactive Output}
To mimic human interaction patterns, ReaperAI utilizes a non-blocking read operation in a separate thread, allowing it to continuously monitor the output as it becomes available for interactive programs. This method involves periodically reading every line of output within a specified time frame, much like a human will wait for, and read command outputs intermittently. The collected data is then updated and fed back to the LLM for further analysis, ensuring that the AI has the most current information to base its next set of commands on. This approach not only enhances the responsiveness of ReaperAI, but also ensures that the AI’s suggestions are grounded in the most recent system responses.

\section{Progress and State Management:}
\subsection{Continuous Evaluation}
ReaperAI implements a robust system of continuous evaluation to monitor the progress of tasks against predefined criteria such as time spent, information gained, and task completion status. Evaluation occurs through a prompting technique by providing both older information and new information, as well as start times and time limits, ReaperAI prompts the LLM to make decisions based on this compiled data. The system uses the concept of diminishing returns to ensure that new information results from recent actions taken by the LLM and to break when that level of new information is not fruitful. This fruitfulness is decided by the LLM by processing the prompt and generating some output. This ongoing assessment is crucial, as it allows for real-time decision-making about whether to proceed with new tasks or refine existing ones. By evaluating each step of the process, ReaperAI ensures that objectives are met efficiently and effectively. This methodology not only maximizes productivity but also enhances the quality of the penetration testing by ensuring that all actions are aligned with the set goals and contribute effectively toward the overarching security objectives.

\subsection{State Updates and Logging}
To maintain a high level of operational integrity, every action taken by the ReaperAI and its outcomes are logged, and the system state is updated accordingly. This comprehensive logging mechanism serves multiple purposes: it ensures traceability of actions, which is essential for debugging and auditing purposes; it enhances accountability by providing a detailed record of operations; and it supports informed decision-making by preserving a historical context of the target system’s state. These updates and logs are instrumental in creating a transparent and effective workflow, where past actions inform future decisions, thereby optimizing the overall penetration testing process. This approach not only improves security assessments but also builds a foundation for more advanced analytics and learning from the accumulated data for the LLM.

\section{Error Handling and Adaptability}
\subsection{Error Handling}

Actual error handling is not fully implemented in the \textit{ReaperAI} proof of concept. For now, the critical error handling in \textit{ReaperAI} involves capturing errors that are piped to \textit{stderr}. These errors occur when commands are executed through the \textit{subprocess} module in Python and are essential for informing the LLM about what transpired during the execution of the commands. This process grants the LLM complete transparency regarding command execution, which is crucial for response adaptation. To transition \textit{ReaperAI} from a proof of concept to a fully functional system, comprehensive error handling will be vital, ensuring robustness and reliability in real-world applications.

\section{Security and Scope Consideration}
\subsection{Constraint Awareness}

\textit{ReaperAI} is engineered with specific constraints to ensure that the penetration testing process remains ethical and minimally disruptive, focusing solely on designated machines. These constraints are pivotal in mitigating potential adverse side effects, such as service disruptions or compromised data integrity on non-target systems. Traditionally, adherence to these constraints is the responsibility of the penetration tester, but integrating these directly into the program has proven effective. By incorporating explicit constraint statements into the prompts, \textit{ReaperAI}'s behavior is modified to consistently adhere to these limits. This approach was particularly verified during initial tests where the generated objectives, such as scanning non-target devices or conducting unauthorized port scans, initially fell outside the intended scope. While these constraints in \textit{ReaperAI} are currently limited, they demonstrate a viable proof of concept for how to develop and enforce limitations to keep AI operations within predetermined boundaries.

Moreover, these constraints systematically guide the behavior of the testing agents, preventing them from making inappropriate assumptions or executing actions that could potentially damage the network or systems. This structured approach not only enhances the precision and effectiveness of the penetration testing process but also supports the ethical standards of cybersecurity practices. By adhering to these established parameters, \textit{ReaperAI} fosters trust and accountability in automated security assessments, ensuring that all activities are ethically sound and justifiable within the research framework.
  
\chapter{Results \& Discussion}

The innovative design of an AI offensive agent, particularly one that integrates a Large Language Model  with advanced retrieval and command execution capabilities, can significantly enhance defensive cybersecurity strategies. By simulating offensive tactics, this agent can uncover vulnerabilities, refine defensive mechanisms, and improve overall security posture. Below are details on the application and use cases of this AI offensive agent in cybersecurity efforts.

\section{Automated Penetration Testing}

\textit{ReaperAI} represents a foundational effort in applying AI to cybersecurity, serving as an initial proof of concept based on preliminary research. While its methodologies and intentions are still in the nascent stages, the project has been intentionally simplified to demonstrate minimal viable functionality. Despite its early developmental stage, \textit{ReaperAI} has achieved notable successes. For instance, it autonomously and successfully exploited the "Eternal Blue" vulnerability on the "Blue" machine on Hack The Box using a metasploit module. This was accomplished using a straightforward one-liner bash command, following the penetration testing methodology outlined previously. The runs are logged in the GitHub project under \textit{src/bashlogs/} due to the verbosity of runs. 

The program is able to successfully run full recon workflows using tools like \textit{ping}, \textit{nmap}, \textit{dns}, and nikto to produce very realistic and usable recon information. The other success is the analyzation and summaries that are accurate and can be used by penetration testers to focus on more important concepts that require more skills. 

Furthermore, \textit{ReaperAI} demonstrated its potential by correctly identifying the exploit for the "Lame" machine on Hack The Box. However, it failed to fully capitalize on this exploit due to an incorrect Metasploit command as well as not having the capability to start multiple threads to 1) start a listener session and 2) execute the exploit found on \textit{Exploit-db}. This highlights both the capabilities and current limitations of \textit{ReaperAI} and showcases the limitation on operating the bash terminal like a human would. These instances illustrate the practical application of the AI in real-world scenarios, providing valuable insights into the effectiveness of its current algorithms and indicating areas for further refinement and development.

In this research, the development of a benchmarking framework was not pursued; instead, the focus was placed on analyzing existing behaviors, which provided more substantive support for the research objectives at hand. This analysis was crucial for understanding the capabilities and limitations of current methodologies, thereby setting a solid foundation for future enhancements. The decision to forego the immediate development of a benchmark framework was based on the need to prioritize in-depth behavioral analysis over establishing performance metrics at this stage.

However, recognizing the importance of benchmarking in assessing the performance and efficiency of offensive security agents, this element is earmarked for future exploration. The inclusion of a benchmarking framework in subsequent research will be critical as AI-driven offensive agents become more sophisticated and widely implemented. This future direction will aim to develop a comprehensive set of standards and metrics that can rigorously evaluate the performance of these agents, ensuring they meet the necessary criteria for effectiveness and efficiency in real-world scenarios. This strategic approach aligns with the overarching goal of advancing the field while ensuring that future developments are measurable and aligned with industry standards.

\section{Unsuccessful Attempts}
ReaperAI experienced multiple instances of interruptions due to various factors. There are numerous points within the agent chain where misconceptions or redundant task progressions can occur. These areas require further refinement to enhance the program's effectiveness and consistency. Some instances where interruptions occurred include:
\begin{itemize}
    \item \textit{sudo} permissions 
    \begin{itemize}
        \item Asking for sudo permissions to conduct elevated scans, but sudo is interactive prompting for a user   
            password
    \end{itemize}
    \item \textit{nc} commands
    \begin{itemize}
        \item Netcat is a network tool to communicate to services through ports, but is vague and interactive
    \end{itemize}
    \item \textit{smbclient} commands because pexepects were not properly done
    \begin{itemize}
        \item SMBClient is a smb shares tool used to connect to smb servers, but interactive command execution 
            wasn't generating the proper expect regex to match the user input prompt
    \end{itemize}
    \item Out of scope on different machines
    \begin{itemize}
        \item The agent started to do host discovery with \textit{nmap} and started to scan other machines on the same vlan
    \end{itemize}
    \item \textit{curl} commands
    \begin{itemize}
        \item Conducting curl commands that were out of scope and assumed the server had a web server
    \end{itemize}
    \item Tried to install applications but got stuck on \textit{sudo} privileges
    \begin{itemize}
        \item The agent tried to \textit{sudo apt install nikto} but didn't have \textit{sudo} privileges
    \end{itemize}
    \item Running \textit{tcpdump} which was out of for the context of the test
    \begin{itemize}
        \item The agent ran a tcpdump command that was out of scope of the test and not within its capabilities.
    \end{itemize}
    \item Assuming information based on "pentesting methodology"
    \begin{itemize}
        \item It would assume information when creating objectives and tasks
        \item  e.g., "Identify SQL injections" when it's unknown if a web server is present
    \end{itemize}
\end{itemize}

During the development of the model, significant challenges were encountered, particularly in the initial stages of transitioning from concept to implementation. Prior to conducting thorough research and following a preliminary literature review, there was an attempt to simply expand the capabilities of an existing program, dubbed \textit{hackingbuddyGPT}, from focusing solely on privilege escalation to encompassing comprehensive penetration testing tasks. This expansion proved problematic, as the program frequently struggled with the assigned tasks, veering into irrelevant tangents and rabbit holes. This lack of focus and direction not only hindered progress but also highlighted the need for a more structured and research-driven approach.

Consequently, these initial setbacks served as a catalyst for more extensive research. The difficulties faced underscored the complexities of adapting AI models to the nuanced and dynamic field of cybersecurity, particularly in the realm of penetration testing. This led to a deeper exploration of the underlying principles and methodologies that could better support such a transition. The subsequent research aimed to refine the model's approach, enhance its task-specific performance, and ensure that its outputs were relevant and practical for real-world cybersecurity challenges. This phase of development was crucial in establishing what is needed to move towards a more robust and effective AI-driven cybersecurity solution.

\chapter{Implementation Challenges, Ethical Considerations, Future Direction}

Exploring the implementation challenges, ethical considerations, and future directions of deploying AI offensive agents in cybersecurity offers a nuanced understanding of the potential impacts and responsibilities associated with this innovative approach. These crucial aspects mentioned above are detailed below.

\section{Challenges in Implementing AI Offensive Agents}

The implementation of AI offensive agents in cybersecurity faces several technical and operational challenges. One primary concern is the accuracy and reliability of the agents' actions, especially in complex and dynamic digital environments. Unfruitful runs, can lead to unnecessary disruptions and resource allocation issues. Additionally, the scalability of AI systems to handle large-scale networks and rapidly evolving threats without compromising performance remains a technical hurdle as well as relying on a closed source LLMs like GPT. There's also the challenge of integrating these advanced AI capabilities with existing cybersecurity infrastructure, requiring significant customization and adaptation to ensure compatibility and effectiveness, as well as adaptiveness for unique tools and unique vulnerabilities. The challenges that an AI offensive agent will face to actually see a significant impact on the domain of penetration testing is great, but for now they are just challenges.

\section{Ethical Considerations in Offensive Cybersecurity}

The use of AI for offensive purposes in cybersecurity introduces a range of ethical considerations that must be meticulously addressed. Key among these is the potential for misuse, where powerful AI capabilities could be leveraged by malicious actors if not properly secured. The development and deployment of AI offensive agents must be guided by strict ethical standards to prevent unintended consequences, such as privacy violations or collateral damage to unintended targets. Moreover, the transparency of AI decisions and actions is crucial to maintain trust and accountability, especially when those decisions may have significant repercussions. In addition to keeping the data secured for each run, the operator would have to have their own collection of capable models to ensure confidentiality, integrity and availability of the data.

\section{Future Directions in AI-Driven Offensive Security}

Looking ahead, the domain of AI-driven offensive security is set to undergo significant advancements, driven by continual improvements in AI technologies and methodologies. This study represents an important step towards shaping the future of this field by highlighting specific areas that require substantial development to yield impactful results:

\begin{enumerate}
    \item Enhanced Command-Line Interaction: One of the primary areas for development is improving the way programming languages, such as \textit{Python}, interact with command-line interfaces to execute interactive commands. While the use of tools like \textit{Pexpect} is a promising development, further research is needed to enable large language models  to effectively run operating systems and utilize various tools. Additionally, the creation of tool-specific models could enhance the functionality and task fulfillment capabilities of these systems.
    \item Expansion of Greater Context Windows: To tackle the broad and complex problems inherent in penetration testing, LLMs require larger context windows. This expansion would allow the models to retain and process more extensive data from previous interactions, enhancing their ability to understand and solve complex security challenges.
    \item Standardization of LLM Outputs: Establishing a standard for processing and utilizing outputs from LLMs is critical. This could involve developing data extraction agents or integrating storage functionalities within LLMs themselves, facilitating easier parsing and application of model outputs within software environments like Python.
    \item Cybersecurity-Specific Embeddings: Introducing domain-specific embeddings for the cybersecurity field could significantly enhance the effectiveness of retrieval-augmented generation (RAG) systems and vector database functionalities. This would allow LLMs to better understand and respond to cybersecurity-specific queries and challenges.
    \item Simulation of Human-Like Abilities: Further research could also focus on simulating more human-like cognitive abilities within LLMs, such as risk analysis and the understanding of fear, pushing the ethical boundaries of AI capabilities. This could enhance the decision-making processes of AI systems in complex and uncertain environments.
    \item Advancement of Task Management: Advancing the development of task generation and completion that mimics human decision-making is another crucial area. This involves creating models that can not only generate and manage tasks but also dynamically adjust their strategies based on changing conditions and priorities, much like a human operator.
    \item Development of Pentesting Benchmarks: Establishing benchmarks for penetration testing will provide a standardized framework to evaluate the efficacy of penetration tests, setting the bar for what constitutes a successful and thorough penetration test.
    \item Integration of Reinforcement Learning: Introducing reinforcement learning tools to better plan an attack can be beneficial to future directions, such ideas include attack graphs or attack maps.

\end{enumerate}
By addressing these areas, future research can significantly advance the capabilities of AI in offensive security, leading to more sophisticated, autonomous, and effective security solutions.

\chapter{Conclusion}
The integration of Artificial Intelligence into offensive cybersecurity represents a transformative shift towards more dynamic, intelligent, and offensive and defense mechanisms. This paper discusses the development of an AI-driven offensive agent, encapsulated within a Python wrapper around a Large Language Model, and enhanced with features such as Retrieval Augmented Generation (RAG), contextual memory, and advanced prompting capabilities. The agent, equipped to simulate cyber-attacks and thereby identify vulnerabilities, also serves to enhance defensive strategies by learning from each interaction.

This exploration delves into the foundational technologies and methodologies that drive the agent’s functionality. These include enhanced prompting, decision-making processes, natural language processing, retrieval augmented generation, task-driven methodologies, AI-driven command generation, command execution, progress and state management, error handling, and security enhancements through constraint awareness. Collectively, these elements enable the AI agent to identify patterns, make informed decisions, and execute complex cybersecurity tasks, culminating in the development of a fully autonomous proof of concept. \textit{ReaperAI}, demonstrates the potential for very effective and dangerous programs to be developed with little effort and understanding of LLMs. The methodologies and techniques provided in this paper also were fully discovered and researched in the lens of offensive security, which at the time of this paper is an untouched topic.

Nonetheless, realizing the full potential of AI in offensive cybersecurity presents substantial challenges and ethical considerations. Technical obstacles such as improving command-line interactions, expanding context windows, standardizing LLM outputs, developing cybersecurity-specific embedding, incorporating reinforcement learning, simulating human-like abilities, enhancing task management, and establishing penetration testing benchmarks must be addressed. Ethically, the development and deployment of such agents must be managed with great care to avoid misuse and ensure transparency and accountability.

In conclusion, leveraging AI within offensive cybersecurity showcases a significant advancement in technology and the domain of cybersecurity. By continuing to innovate and address challenges with cutting edge technology, cyber defense teams can stay on top of upcoming threats and foster a more secure and resilient digital environment.

%================================================================
% References section.  
%\bibliographystyle{apalike}
\bibliographystyle{plainnat}
\bibliography{references}

%================================================================
% Appendices section.  
% If you don’t have appendices, comment out these lines
   \appendix
   \chapter{Prompts}
   \section{query\_next\_command.txt}
\begin{verbatim}
Your current role:
"${current_role}"

You can only: 
    - give a command to be executed on the shell that is NOT
    interactive. 
    You don't have that integration yet and I will respond with the
    terminal output when running this c
    ommand on the linux server. Only state the
    command to be executed. 
    - give commands that do not contain `nc`

Your current task is:
"${current_task}"

Your current constraints are:
"${constraints}"

Your current target is:
"${target.ip}"

% if len(history) != 0:
You already tried the following commands and their arguments:

'''bash
${history}
'''

Do not repeat already tried attacks.
%endif

You currently know the following about the target system:
"${state}"

Based on this you have thought about it and the analyzation you have 
concluded on from the previous step is: "${analyzation}"

% if task_tree:
Your task tree to keep track of this pentest:

${task_tree.generate_task_summary()}
%endif

Given the current role, objective, constraints, history, and state, and
analysis of Give your command that will best accomplish the task. 
Do not add any explanation or add an initial `$`
\end{verbatim}

\section{analyze\_cmd.txt}
\begin{verbatim}
As a penetration tester analyzing the outputs of a command, 
I've executed'${cmd}' against the system's defenses. 
I'll share the stdout, stderr, and exit code with you. 
Your task is to distill this data into a concise summary,
capped at ${summary_max_chars} characters. Focus your 
analysis uncovering insights related to security 
vulnerabilities, performance bottlenecks, or any signs 
of instability or unreliability in the system or its 
services. Highlight critical findings that could indicate 
potential risks or weaknesses in the system's architecture 
or configurations and explain step by step

Command: ${cmd}
Stderr: ${stderr}
Stdout: ${resp}
Summary (in ${summary_max_chars} or fewer characters):
\end{verbatim}
\section{evaluate\_progress.txt}
\begin{verbatim}
You are an advanced LLM designed to assist in 
cybersecurity operations, specifically in evaluating 
the progress of "${stage}" tasks during penetration
testing phases. Based on the given inputs, your goal 
is to determine the appropriate next steps in the 
"${stage}" process by considering time limits,
coverage thresholds, and diminishing returns.

Inputs:

List of "${stage}" Tasks: ${task}

Constraints:

Time Limits: Each "${stage}" task has a maximum 
allotted time of ${max_time_per_task} seconds. If 
the cumulative time which is ${cur_time} reaches this 
limit from the start time which is ${start_time}, 
recommend moving to the next phase. 

Diminishing Returns: "${stage}" is limited to a maximum of
${max_recon_iterations} iterations. If additional iterations 
yield less than ${min_new_information} new information, 
indicate that diminishing returns have been reached and recommend
stopping "${stage}" to proceed further. 

Decision Points:
Evaluate if the time limit for a "${stage}" task has 
been exceeded. If so, advise moving to the next phase. 
Determine if the "${stage}" efforts are experiencing 
diminishing returns based on the number of iterations 
and the percentage of new information gained from the new:
${new_information} compared to the old: ${old_information}. 
If diminishing returns are detected, recommend
stopping "${stage}" and proceeding to the next step. 

Output:
Provide a recommendation. The recommendation should 
indicate whether to continue "${stage}", move to the 
the next phase, or conclude the "${stage}" phase due 
to coverage thresholds being met or diminishing returns.
\end{verbatim}

\section{get\_objectives}
\begin{verbatim}
You are a cybersecurity pentester and not a redteamer
conducting a BLACK BOX pentest on one single machine 
which means we know nothing prior about this machine. 
Given the current stage in your penetration testing
cycle named "${stage}", produce a realistic and tangible 
ordered list of 3 most important tasks of the given stage 
that can be completed to finish the ${stage} that does
NOT assume anything about the machine, its ports or 
services and should be solely based on the context 
I give you here: ``` ${background} ``` This list
should encompass tasks that are specific to the stage, 
and can be properly executed and fulfilled to ultimately 
proceed to the next stage. Do NOT give any explanations 
or descriptions. Again, ensure the tasks do NOT assume any
information about the target. Objectives need to be brief 
and very specific.
Output the list in an json array.
\end{verbatim}
%     ...
%  READ PERMISSIONS BEFORE COMMENTING OUT OR REMOVING
% \chapter{PERMISSIONS}
%    This is your last appendix. 
   
%    Place all of your permissions here for everything used from copyrighted material: i.e., long quotations, tables, and figures. Label each permission page with the list of figure numbers or tables or sections of your text that is from the copyrighted material.
   
%    Most journals have an online permissions process where you submit what you are using from the journal and an automatic system sends you permissions for a thesis/dissertation OR they have a blanket permission for use in a thesis/dissertation. 
   
%    You must have permission for your own published work when you have signed a copyright release form; the above description for journals applies here as well. 
   
%    If you are using material from an NMT thesis or dissertation, you may use the permission page at the end of the referenced thesis as your permission. 
%     ...

% 
\copyrightpage

\end{document}